# Thermal stability of Te-hyperdoped Si: Atomic-scale correlation of the structural, electrical and optical properties

Mao Wang[1,2,*], R. Hübner,[1] Chi Xu[1,2], Yufang Xie[1,2], Y. Berencén[1], R. Heller[1], L. Rebohle[1], M. Helm[1,2], S. Prucnal[1], and Shengqiang Zhou[1]

[1]Helmholtz-Zentrum Dresden-Rossendorf, Institute of Ion Beam Physics and Materials Research, Bautzner Landstr. 400, 01328 Dresden, Germany

[2]Technische Universität Dresden, 01062 Dresden, Germany

**Abstract**

Si hyperdoped with chalcogens (S, Se, Te) is well-known to possess unique properties such as an insulator-to-metal transition and a room-temperature sub-bandgap absorption. These properties are expected to be sensitive to a post-synthesis thermal annealing, since hyperdoped Si is a thermodynamically metastable material. Thermal stability of the as-fabricated hyperdoped Si is of great importance for the device fabrication process involving temperature-dependent steps like ohmic contact formation. Here, we report on the thermal stability of the as-fabricated Te-hyperdoped Si subjected to isochronal furnace anneals from 250 °C to 1200 °C. We demonstrate that Te-hyperdoped Si exhibits thermal stability up to 400 °C with a duration of 10 minutes that even helps to further improve the crystalline quality, the electrical activation of Te dopants and the room-temperature sub-band gap absorption. At higher temperatures, however, Te atoms are found to move out from the substitutional sites with a migration energy of $E_M = 2.1 \pm 0.1$ eV forming inactive clusters and precipitates that impair the structural, electrical and optical properties. These results provide further insight into the underlying physical state transformation of Te dopants in a metastable compositional regime caused by post-synthesis thermal annealing as well as pave the way for the fabrication of advanced hyperdoped Si-based devices.

[*]Corresponding author, email: <u>m.wang@hzdr.de.</u>

# I. INTRODUCTION

The nonequilibrium method of ion-implantation followed by nanosecond-pulsed laser melting (PLM) is extensively used for doping silicon with impurity concentrations several orders of magnitude above the solid solubility limit (hyperdoping) [1-3]. Such a hyperdoping approach has attracted interest for decades due to the possibility of designing Si with unique and exotic physical properties. In the microelectronic industry, Si hyperdoped with shallow-level impurities such as As (or P, B, Sb) allows for the formation of shallow junctions with high electrical conductivity [3-5]. Recently, it has been shown that Si hyperdoped with deep-level (with a much higher ionization energy) impurities (Ti, Ag, Au, S, Se and Te) exhibits strong sub-bandgap absorption caused by the presence of an intermediate band within the band gap of Si. This finding is a major step toward the implementation of room-temperature broad band infrared photodetectors [6-14] and high-efficiency solar cells [15-20]. Moreover, hyperdoped Si provides a good prospect to explore the impurity-mediated insulator-to-metal transition (IMT) in Si [21-25].

However, it should be noted that hyperdoped Si is a metastable and supersaturated solid solution system. The novel physical properties mentioned above are expected to be strongly altered by a subsequent thermal treatment [26-31]. Therefore, a comprehensive study regarding subsequent thermal treatments is of great interest not only for the nature of thermal deactivation, but also for the thermal stability of the necessary subsequent processes to produce the ultimate devices by complementary-metal-oxide-semiconductor (CMOS)-compatible processing. Previous investigations on hyperdoped Si with shallow-level dopants suggested that the main mechanism behind the thermally-instable deactivation is the formation of defect complexes (dopant-vacancy clusters or precipitates) depending on the thermal anneals [30,32,33]. The thermal deactivation of sub-bandgap absorption in hyperdoped Si with deep-level impurities is attributed to the long-range dopants diffusion (in polycrystalline Si) or to the formation of dopant-defect complexes (in single-crystalline Si) [29,31]. However, a direct correlation between the impurity lattice location and the electrical and optical properties upon thermal annealing has yet to be established.

In this work, we explore the thermal stability of epitaxial Te-hyperdoped Si layer by using furnace annealing in the temperature range of 250-1200 °C. We demonstrate that the as-fabricated Te-hyperdoped Si layer possesses a thermal stability upon annealing for 10 minutes up to 400 °C. In particular, the crystalline quality, the Te substitutionality, the electrical activation and the mid-infrared absorption slightly improve, which is crucial for producing CMOS-compatible devices.

Moreover, we show that the deactivation process at annealing temperatures higher than 500 °C results from the thermally-mediated atomic-movements of Te dopants to form inactive complexes such as clusters and precipitates in the metastable hyperdoped system. This finding provides further insight into the underlying physical state transformation of Te dopants in a metastable compositional regime caused by thermal treatment.

## II. Methods

### A. Sample preparation

Two kinds of (100) Si wafers were used for the ion implantation process: (a) single side polished intrinsic Si wafer ($\rho > 10^4$ Ω·cm) for structural and electrical characterization, and (b) double side polished $p$-type Si wafer (boron-doped, $\rho \approx 1–10$ Ω·cm) for the investigation of the optical properties. Both the implantation and the annealing conditions were always identical. Samples were prepared by a combined implantation of Te ions with implantation energies of 150 keV and 50 keV (to obtain a homogeneous distribution of Te in the implanted layer) at room temperature to a dose of $4.7 \times 10^{15}$ cm$^{-2}$ and $1.9 \times 10^{15}$ cm$^{-2}$, respectively. Samples were tilted by ~7° off the incident beam to minimize channeling effect. Based on the previous Rutherford backscattering-channeling spectrometry (RBS/C) measurements and transmission electron microscopy (TEM) results [12], the implanted layer has a thickness of 120 nm with a peak Te concentration of $7.5 \times 10^{20}$ cm$^{-3}$. The ion-implanted samples were subsequently irradiated in air with one pulse by a spatially homogenized, pulsed XeCl excimer laser (Coherent COMPexPRO201, wavelength of 308 nm, pulse duration of 28 ns) with a square spot of approximately 5 mm × 5 mm [12]. At this laser density, the Te-implanted amorphous Si layer recrystallizes via liquid-phase epitaxial growth with a solidification speed in the order of 10 m/s while cooling down [34]. After PLM, the recrystallized layer is single-crystalline, free of extended defects, and doped with a Te concentration of around 1.5 at. % [12]. After PLM, all samples were cleaned by HF acid to remove the native SiO$_2$ layer and thereafter are referred as 'as-fabricated' samples in this manuscript. Then the as-fabricated samples were individually subjected to thermal annealing in a furnace (Ströhlein instruments, Germany). For the thermal stability analysis, isochronal anneals were performed covering a temperature range of 250 °C to 1200 °C with a duration of 10 minutes in a nitrogen atmosphere (purity: 99.99%).

### B. Characterization

RBS/C was carried out to characterize the structural properties of samples before and after furnace annealing. The RBS/C measurements were performed using a 1 mm diameter collimated 1.7 MeV He$^+$ beam of the Rossendorf van de Graff accelerator with a 10-20 nA beam current. The backscattered particles were detected at an angle of 170° with respect to the incoming beam using silicon surface barrier detectors with an energy resolution of 15 keV. The channeling spectra were collected by aligning the sample to make the impinging He$^+$ beam parallel to the Si <100> axes. Also, angular axial scans around the <100> axis were measured using a two-axis goniometer. The integral backscattering yields in energy windows corresponding to scatterings from Si and Te atoms are then plotted as a function of the tilt angle. High-resolution TEM was performed on an image C$_s$-corrected Titan 80-300 microscope (FEI) operated at an accelerating voltage of 300 kV to analyze the microstructure of the samples before and after furnace annealing. To carry out qualitative chemical analysis, high-angle annular dark-field scanning transmission electron microscopy (HAADF-STEM) imaging and spectrum imaging based on energy-dispersive X-ray spectroscopy (EDXS) were done at 200 kV with a Talos F200X microscope equipped with an X-FEG electron source and a Super-X EDXS detector system (FEI). Prior to STEM analyses, the specimen mounted in a high-visibility low-background holder were placed for 10 s into a Model 1020 Plasma Cleaner (Fischione) to remove possible contamination. Classical cross-sectional TEM lamella preparation was done by sawing, grinding, polishing, dimpling, and final Ar ion milling.

The electrical properties of the Te-hyperdoped Si layers before and after subsequent furnace annealing were measured by a commercial Lakeshore Hall measurement system in the van der Pauw configuration [35] under a magnetic field perpendicular to the sample plane. The magnetic field was swept from -5 T to 5 T. The gold electrodes were firstly sputtered onto the four corners of the square-like samples. Prior to the sputtering process, the native SiO$_2$ layer was removed by etching in hydrofluoric acid (HF). Then, a silver conductive glue paste was applied to contact the wires to the gold electrodes. All contacts were confirmed to be ohmic by measuring the current-voltage curves at different temperatures.

The optical properties were investigated by Fourier transform infrared spectroscopy (FTIR) using a Bruker Vertex 80v FT-IR spectrometer. In detail, transmittance (*T*) and reflectance (*R*) measurements were performed at room temperature with a KBr beam-splitter and a mid-IR DLaTGS detector in the infrared spectral range of 0.05-0.85 eV ($\lambda$ = 1.4-25 μm). The sub-bandgap absorptance (*A* = 1−*T*−*R*) was then determined by recording the transmittance and reflectance spectra. A gold mirror was used as a 100% reflectance standard, and the illumination area of the spectrometer was focused on a spot diameter of about 3 mm to ensure that the measurements were solely performed within the sample area (5 × 5 mm$^2$).

## III. RESULTS AND DISCUSSION

## A. Thermal stability of the structural properties

Figure 1(a) depicts the representative RBS spectra in random/channeling configuration of Te-hyperdoped Si layers after furnace annealing. The reference channeling spectra of the as-implanted sample, the as-fabricated sample and the bare silicon substrate are also included. As being well established, $\chi_{min}$ (channeling minimum yield), which is defined as the ratio of the backscattering yield of the channeling to the random spectra, represents the degree of lattice damage in the host material. Consistent with previous works [12,36], the channeling spectrum of the as-implanted Te-hyperdoped Si layer shows a broad damage peak from 870 keV to 935 keV with a $\chi_{min}$ of 94%, which indicates that the as-implanted layer is completely damaged and amorphous. On the contrary, the backscattering yields in channeling condition of the as-fabricated sample has a $\chi_{min}$ of 4% (from 800 keV to 910 keV), which points out that a high-quality single crystal has been restored after the PLM process [12]. Unexpectedly, the crystalline quality of the furnace annealed samples with annealing temperatures below 500 °C remains good because the $\chi_{min}$ values are comparable to that of the as-fabricated sample. However, for samples with furnace annealing temperature above 500 °C, a clear increase of $\chi_{min}$ was observed. This suggests an increased amount of disorder and a low crystalline quality in the Te-hyperdoped layer which were induced by high-temperature furnace annealing.

Figure 1(b)-(f) show the RBS/C Te signal taken from both the random and channeling directions of the selected Te-hyperdoped Si layers juxtaposed with the corresponding spectra of the as-implanted layer. As shown in FIG. 1(b) and (c), upon annealing up to 400 °C the depth profile of Te is not significantly changed, but the channeling spectrum yield is reduced near the surface region (around 1490 keV). This proves that the Te dopants remain at the Si lattice sites and part of the interstitials are moved to the substitutional Si sites [37], particularly near the surface upon annealing up to 400 °C. However, the channeling effect becomes deteriorated starting from the deeper depth of the doped layer as the furnace annealing temperature increases, as shown in FIG. 1(d) and (e). Finally, the RBS random and channeling spectra of Te signal are nearly

overlapped in sample-1200 °C (see FIG. 1(f)), where an inner-diffusion of Te can be clearly observed as compared to the as-implanted sample.

To gain a quantitative estimation, the Te substitutional fraction is approximated by:

$$f_{sub} = \frac{1-\chi_{min}(Te)}{1-\chi_{min}(Si)} \quad (1)$$

where $f_{sub}$ is the fraction of impurity (Te) in substitutional site, $\chi_{min}(Te)$ is the ratio of the channeling and random Te yield integrated over the recrystallization depth and $\chi_{min}(Si)$ is the corresponding value for Si at the same depth along the [100] direction [37]. As shown in FIG. 1(g), the substitutional fraction of Te in furnace annealed samples with annealing temperature up to 400 °C exhibits a slight increase. Afterwards, it decreases as the temperature increases above 400 °C. Therefore, this suggests that the structural stability of Te-hyperdoped Si layer can be preserved up to a post-thermal anneal of 400 °C for 10 minutes.

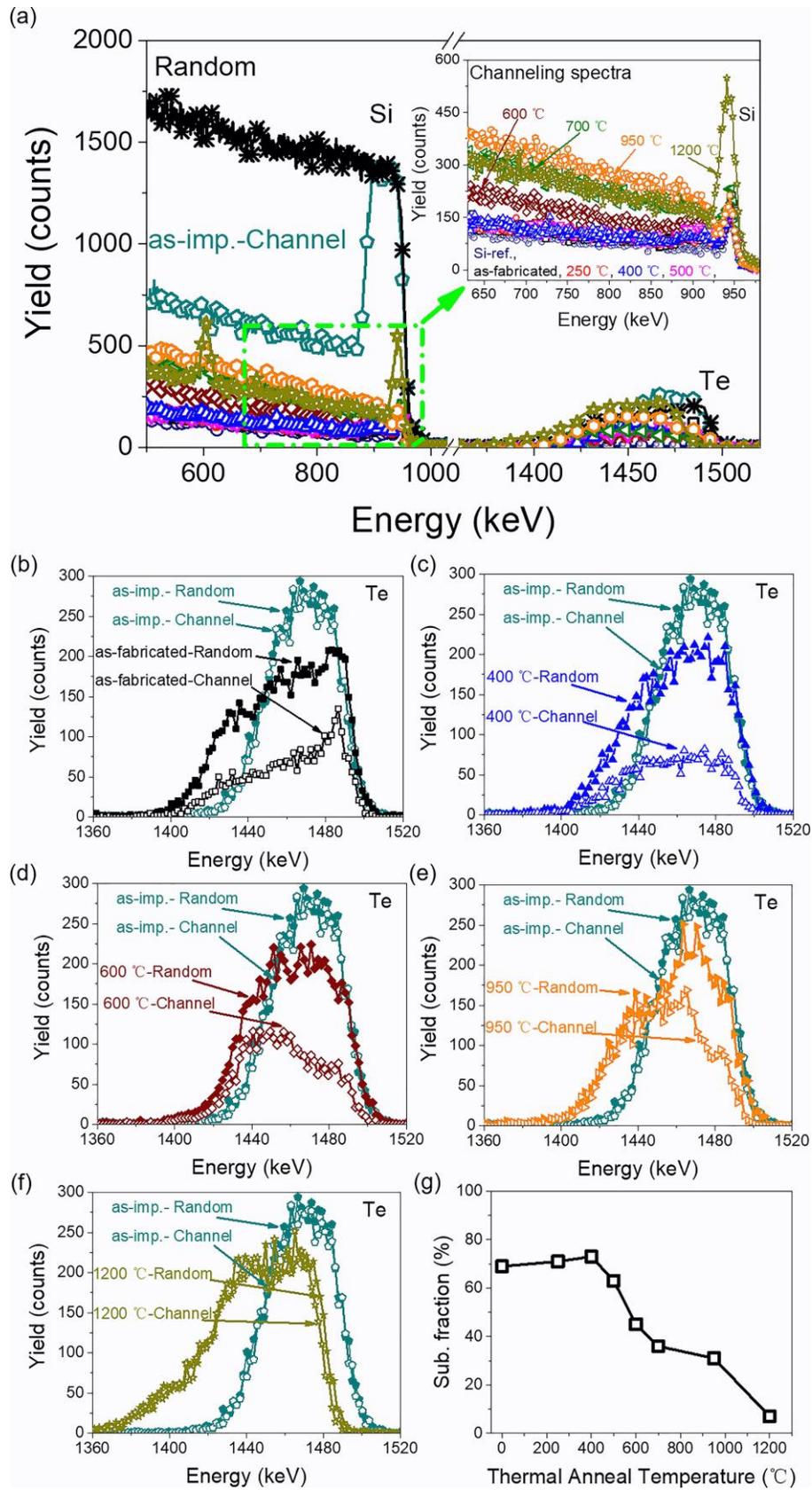

FIG. 1 (a) A sequence of RBS/C spectra of Te-hyperdoped Si layers furnace annealed for 10 minutes at different temperatures (red: 250 °C; blue: 400 °C; magenta: 500 °C; wine: 600 °C; olive: 700 °C; orange: 950 °C; dark yellow: 1200 °C). The channeling spectrum of the as-fabricated sample (black) and the virgin Si (navy) are also included as references. The inset shows the magnification of Si signals from the channeling spectra. (b)-(f) The corresponding random and channeling Te profiles compared to the as implanted Te distribution measured by RBS/C ((b) the as-fabricated sample, (c) 400 °C, (d) 600 °C, (e) 950 °C and (f) 1200 °C). (g) The Te substitutional fraction as a function of the furnace annealing temperature.

Figure 2 shows representative HAADF-STEM images and corresponding HRTEM micrographs of selected Te-hyperdoped Si layers. For the as-fabricated sample, single-crystalline regrowth is confirmed by HRTEM imaging (FIG. 2(b), also described previously in ref.[12]), while Te is distributed homogeneously within the top 120 nm (shown in FIG. 2(a)). Furnace annealing at 400 °C does not change the microstructure, i.e. the epitaxial Te-hyperdoped layer is not affected (FIG. 2(d). The Te distribution is laterally homogeneous, but shows a rather smooth transition with depth (FIG. 2(c)). Importantly, none of Te surface segregation, nano-scale Te agglomerates, extended defects, secondary phases or cellular breakdown is detected.

However, Te-related complexes appear in the sample annealed at 500 °C and more pronounced for that annealed at 600 °C. As shown by the appearance of small bright dots in the HAADF-STEM images in FIG. 3(a) and (b), Te starts diffusing and tends to create small clusters at 500 °C and 600 °C, respectively. Hence, the annealing-related energy is high enough to induce the segregation of Te atoms. For both samples, the microstructure of the recrystallized layer remains single-crystalline (not shown here). At even higher furnace annealing temperatures, such as 950 °C, Te can be found at defects/stacking faults which indicates that the Si crystalline quality is no longer uniform. Additionally, the segregated Te clusters grow and form Te-rich secondary phase particles in the doped layer, which have a diameter on the order of 5 nm, some of them reaching a diameter of 20 nm (FIG. 2(e) and (f)). These larger Te-rich clusters are mainly observed in the deeper region of the hyperdoped layer. This is in good agreement with the RBS/C data, where the channeling effect is no longer obvious particularly in the deeper region, and the Te substitutional fraction decreases significantly for the sample-950 °C. For the furnace annealed sample annealed at 1200 °C, the Te-rich clusters become larger and are distributed within the whole layer. Additionally, the silicon oxide surface layer is found to grow.

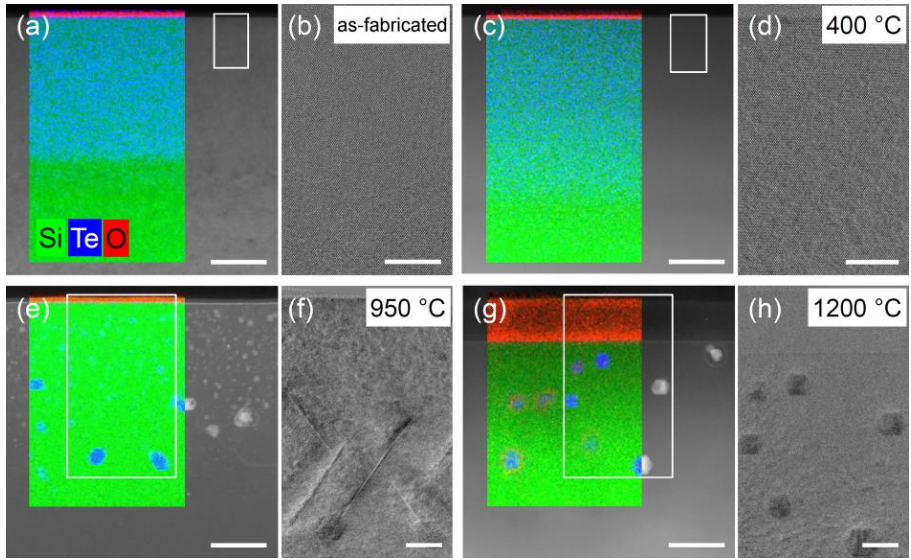

FIG. 2 Cross-sectional HAADF-STEM images superimposed with the corresponding EDXS element maps (blue: tellurium, green: silicon, red: oxygen) and representative HRTEM images for the field of view as depicted by the associated white rectangle) for selected Te-hyperdoped Si layers. (a) and (b) as-fabricated sample; (c) and (d) 400 °C; (e) and (f) 950 °C; (g) and (h) 1200 °C. Scale bars: 50 nm for (a), (c), (e) and (g); 20 nm for (f) and (h); 10 nm for (b) and (d).

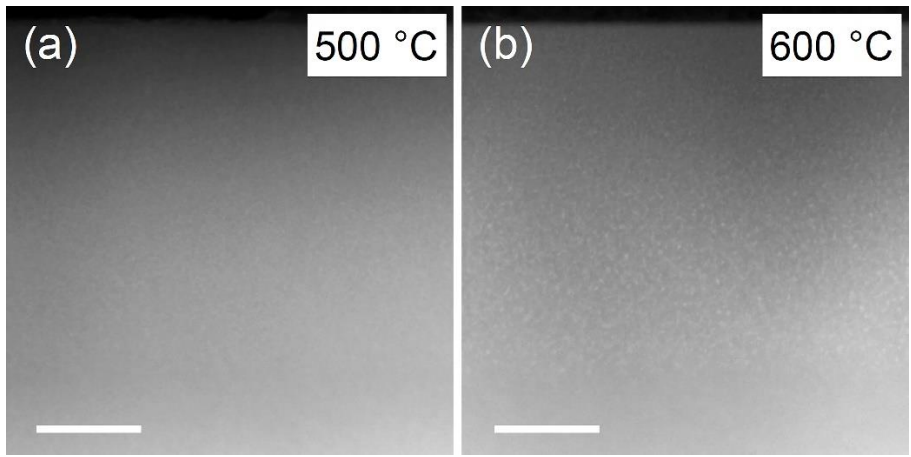

FIG. 3 Cross-sectional HAADF-STEM images for Te-hyperdoped Si layers after furnace treatment at 500 °C (a) and 600 °C (b) for 10 min., respectively. Scale bars: 50 nm.

## B. Evolution of the impurity lattice occupation with annealing

To unambiguously determine the incorporation site of Te in Si subjected to various furnace annealing temperatures, RBS/C full angular scans about the [100] axes were performed. FIG. 4 shows the angular distributions of the normalized yield of He ions backscattered by Te and Si at the same depth as a function of tilt angle for the furnace annealed samples with different temperatures. The corresponding angular scan of the as-fabricated sample is also included as a reference. As described in the previous work [39], the two typical parameters $\chi_{min}$ (the minimum value of the normalized yield in the angular scan, which is equivalent to the previous definition in the section III.A) and $\psi_{1/2}$ (the half width at half maximum of the angular scans between $\chi_{min}$ and 1) are used to characterize the angular distributions [38-40]. $\chi_{min}$-$Te$ is higher than $\chi_{min}$-$Si$ and $\psi_{1/2}$-$Te$ is narrower than $\psi_{1/2}$-$Si$, as shown in FIG. 4(a). As well-established before [38-41], the angular scan of Te will overlap with that of Si in case of substitutional Te ($S_{Te}$). A peak is expected to arise in the middle of the angular scan for interstitial Te ions. The angular scan of substitutional Te atoms but slightly displaced from the Si lattice is expected to be narrow and shallow. Therefore, for furnace annealing temperatures up to 400 °C the majority of Te atoms are located on substitutional sites but with small displacement (see FIG. 4(a), (b) and (c)) which are likely the substitutional $S_{Te}$-$S_{Te}$ dimers [36], and a small amount of Te is at interstitial sites. At annealing temperatures higher than 400 °C, however, $\chi_{min}$-$Te$ gradually increases and $\psi_{1/2}$-$Te$ gradually decreases, as shown in FIG. 4(d)-(h). These results suggest that as the furnace annealing temperature increases, the majority of Te dopants gradually move out of the lattice positions into non-substitutional sites to form $Te$-$V_m$ complexes ($V_m$ denotes $m$ Si vacancies, as reported in As-hyperdoped Si [42]), clusters and even precipitates, which has been confirmed by the TEM results (see FIG. 2(e)-(h)).

The angular scan results show the different lattice sites prevailing at different annealing temperatures, as a result of their thermal stabilities. From FIG. 1(g), one can identify at which temperature the substitutional fraction starts decreasing. Therefore, following the procedure of Arrhenius models for the thermally activated migration described in Refs. 45-47, the migration energy ($E_M$) of $S_{Te}$-$S_{Te}$ can be obtained by:

$$E_M = k_B T ln\left[\frac{\mathcal{V}_0 \Delta t}{N} \frac{1}{ln\left(\frac{f_{n-1}}{f_n}\right)}\right] \quad (2)$$

where $k_B$ is the Boltzmann constant, $T$ the furnace annealing temperature (here, $T$ = 500 °C is applied in the calculation), $\Delta t$ the furnace annealing duration (in seconds), $\mathcal{V}_0$ the attempt frequency, here using a typical value of $10^{12}$ s$^{-1}$, i.e. of the order of the lattice vibrations, $f_{n-1}$ the substitutional fraction before annealing at $T$, $f_n$ the fraction after the annealing at $T$ and $N$ the required number of steps before a substitutional Te atom moves off from $S$ site into a random position. As described before [43-45], a simple estimation for the maximum $E_M$ corresponds to the assumption that only one jump ($N = 1$) is considered enough to move $S_{Te}$ atom out to a random position, while the minimum $E_M$ relates to a reasonable assumption on the maximum number of jumps ($N_{max}$) that take place before the $S_{Te}$ atom is kicked out substitutional site into a random position. Here, the typical value of $N$ is applied by considering most vacancies will annihilate during annealing or form larger multi-atom Te complex. Therefore, a migration energy is found to be $E_M = 2.1 \pm 0.1$ eV, which is higher than that measured for S-hyperdoped Si, around 1.8 eV [29]. This indicates that Te-hyperdoped Si has a better thermal stability.

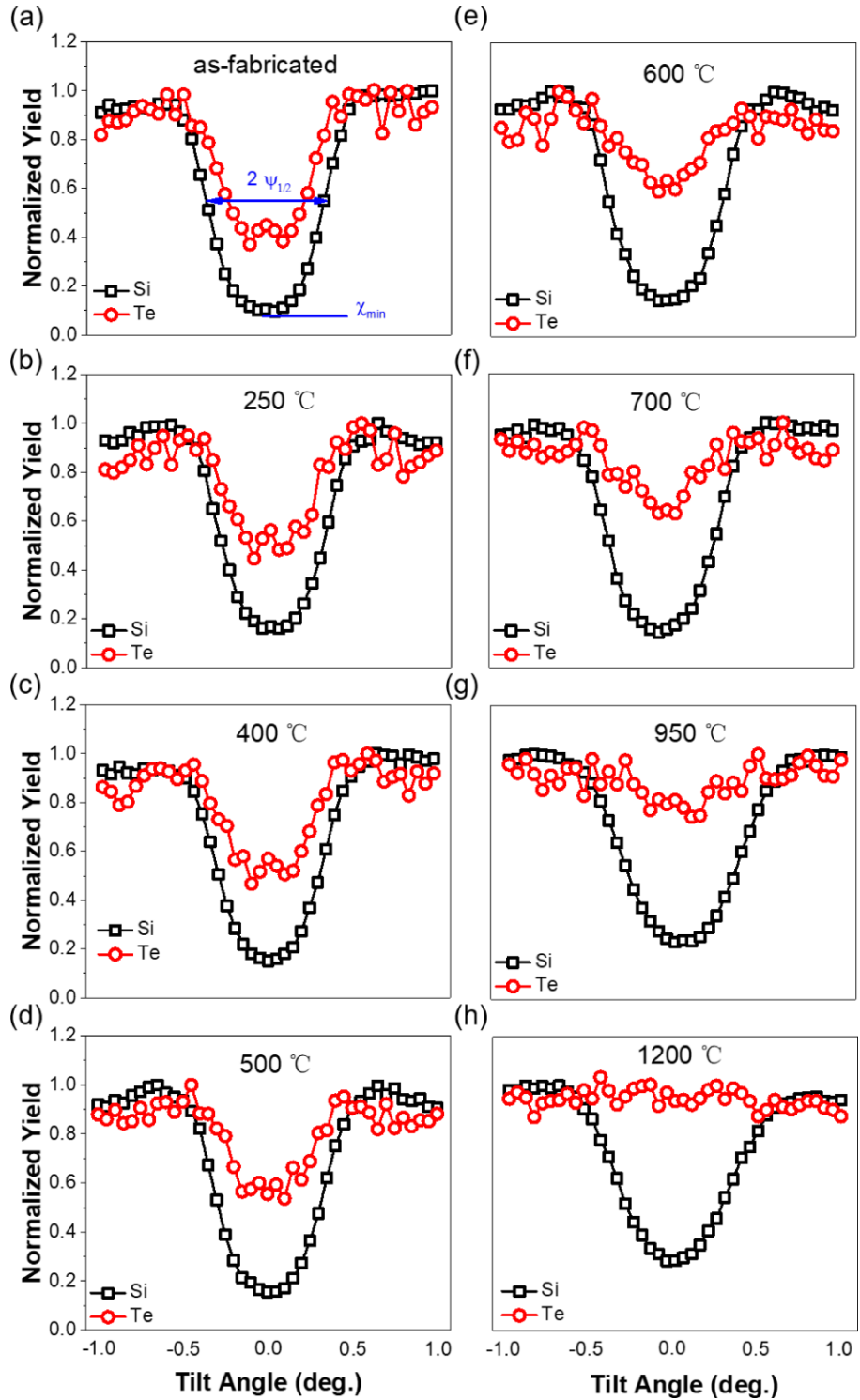

FIG. 4. Angular scans about the <100> axes for Te-hyperdoped Si samples of (a) the as-fabricated-treated sample, (b) - (h) samples with the same condition after furnace annealing with different temperatures for 10 minutes. Angular distributions have been normalized in a random direction. The red circle corresponds to Te atoms and black square to Si. Solid lines are drawn through the symbols.

## C. Thermal stability of the electrical-transport properties

Electrical characterization has been performed to gain further insight into the physical state of the Te dopants in Si. To avoid the influence of the parallel conduction from the Si substrate, samples prepared on the *intrinsic* Si substrate ($\rho > 10^4$ Ω·cm) were measured. FIG. 5(a) shows that the temperature-dependent sheet resistance ($R_s$) of furnace annealed samples evolves with anneal temperature. The as-fabricated sample is also included as a reference. At room temperature, the $R_s$ of the furnace annealed samples is less than 600 Ω/square, which is corresponding to a resistivity of less than $10^{-2}$ Ω·cm (using the effective thickness of 120 nm). This is much lower than that of the intrinsic Si substrate, which confirms that the intrinsic Si substrate has no influence on the transport properties of the furnace annealed layers considering the respective thickness. $R_s$ has a remarkably non-monotonic evolution, which firstly decreases with increasing furnace annealing temperature. However, the samples annealed at 950 °C and 1200 °C show a remarkable increase in resistance and change from metallic-like to semiconductor-like as seen from the temperature-dependent resistance. Moreover, the free electron concentration ($n$) and mobility (µ) are plotted versus the furnace annealing temperature in FIG. 5(b). As expected, $n$ also exhibits a non-monotonic evolution with annealing, which is of the same qualitative trend as $R_s$. In detail, $n$ first increases by 27% (annealing temperature ≤ 400 °C) and then decreases to about 2% (1200 °C) of its initial value (see the black square in FIG. 5(b)). Meanwhile, µ remains essentially constant at around 14 cm$^2$/V·s with annealing temperature ≤ 400 °C and subsequently increases monotonically to 310 cm$^2$/V·s after annealing at 1200 °C.

The remarkable non-monotonic evolution of electrical properties shown here is in agreement with the deactivation of S-hyperdoped Si reported in Ref 33. Here, a possible mechanistic interpretation is proposed to explain the non-monotonic evolution. The initial electrical-active states are substitutional Te dopants (single $S_{Te}$ or $S_{Te}$- $S_{Te}$ dimers), which are the dominant states in Te-hyperdoped Si (verified by the RBS/C results). Then, after annealing at lower temperatures, the increase of $n_c$ is attributed to the increase of the substitutional fraction of Te dopants (see FIG. 1(g)) and the decrease of the residual disorder after the PLM process [46], while the unchanged µ is due to the dominant ionized impurity scattering. Particularly, the annealing between 500 and 700 °C presents an intermediate stage where $n$ decreases but µ increases, leading to a smaller $R_s$. After annealing at even higher temperatures, electrically-inactive complexes (e.g. vacancy-impurity complexes, clusters or precipitates) tend to form. Moreover, the fraction of these inactivated states increases as the annealing temperature increases, which results in the eventual decrease of $n$. This interpretation of the electrical properties by the assignment of the dopant-states induced by the thermally-activated atomic movements is confirmed by the experimental data shown in Sec. **A** and **B**.

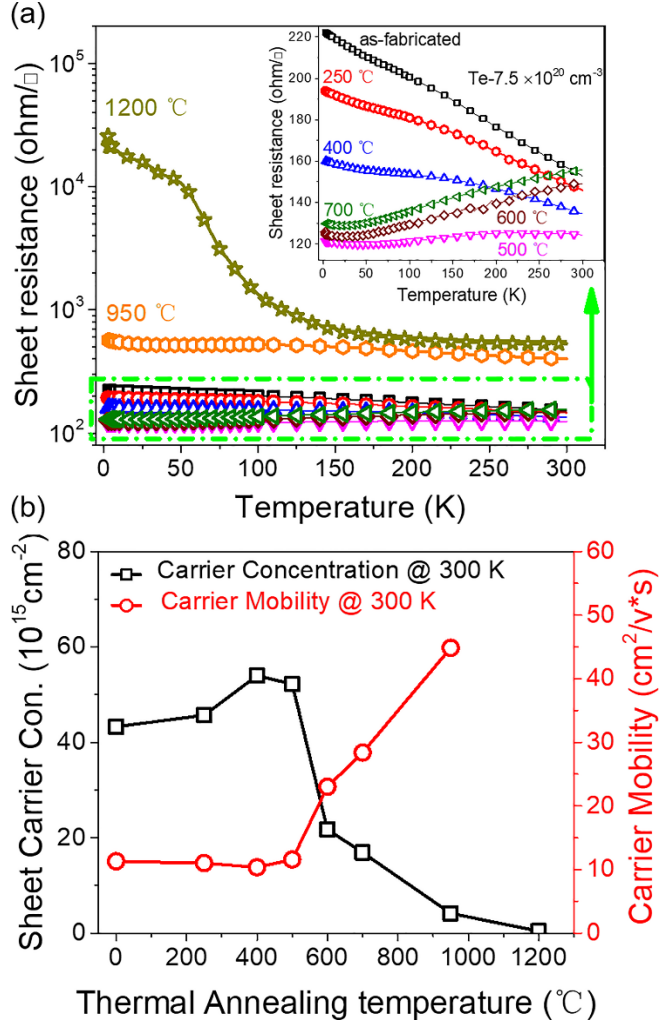

FIG. 5. Evolution of the electrical properties for the furnace annealed Te-hyperdoped Si samples with an annealing duration of 10 min at different annealing temperatures. The annealing temperatures are indicated in the labels. The as-fabricated sample is also included as a reference. (a) Temperature-dependent sheet resistance $R_s$ in the temperature range of 2 K-300 K. The inset shows the magnification of the data for samples with lower annealing temperatures. (b) Sheet carrier concentration $n_c$ (black square) and carrier mobility µ (red circle) for each of the samples in (a). Solid lines drawn through the symbols are a guide for the eye.

## D. Thermal stability of the sub-bandgap optical absorption

The evolution of the sub-bandgap absorptance as a function of furnace annealing temperature was investigated to analyze the optical activity of the Te dopants in Si. Accompanying the formation of Te complexes (inactive centers) and the concomitant decrease in the substitutional fraction (as shown in Sec. **A** and **B**) at higher temperature anneals, the sub-bandgap optical absorptance is also expected to decrease. Figure 6(a) shows the optical absorptance for as-fabricated Te-hyperdoped Si samples subjected to furnace annealing at different temperatures. The as-fabricated sample and the bare Si substrate are also shown as a reference. The absorptance below 0.2 eV can be partially attributed to the free carrier absorption [47]. Particularly, a strong and broad sub-bandgap optical absorptance (from 0.20 eV to 0.87 eV) is observed in the as-fabricated sample and in the furnace annealed samples while the reference silicon substrate has no such absorptance. This is consistent with the previous reported works [12]: the well-defined broad absorptance band comes from the presence of Te-mediated impurity band (IB) in the upper half of the Si bandgap. As shown in FIG. 6(a), the sub-bandgap optical absorptance in the Te-hyperdoped Si presents a pronounced non-monotonic evolution with furnace annealing. The strongest broad-band absorptance is found to take place when a post-furnace annealing at 400 °C for 10 min is applied.

To quantify the sub-bandgap absorptance for each sample, the strength of the optical absorption is defined as the integral of the absorptance spectrum from 0.20 eV to 0.87 eV [12,47]. FIG. 6(b) shows the integrated sub-bandgap absorptance ($A_{Int}$) as a function of furnace annealing temperature, where the as-fabricated sample presents an $A_{Int}$ of 0.31 as the initial value. $A_{Int}$ firstly increases by 21% with increasing annealing temperature up to 400 °C and then gradually decreases to about 0.072 by 78% of its initial value if the annealing temperature increases up to 1200 °C (see FIG. 6(b)). The enhancement of $A_{Int}$ for samples annealed at 250 °C and 400 °C not only corresponds to the increase of the substitutional Te, but can also be partially attributed to the removal of the lattice distortion and the residual disorder after the prior pulsed laser melting process [46]. The apparent reduction of $A_{Int}$ with the furnace annealing temperature increasing from 500 °C to 1200 °C is related to the change in the metastable atomic configuration of Te atoms in the Si matrix, which is confirmed by the RBS results. The non-monotonic evolution of $A_{Int}$ with the thermal anneals is consistent with the variation of the substitutional Te fraction.

Figure 6(c) displays the integrated sub-bandgap absorptance as a function of substitutional Te concentration. A direct correlation is observed between $A_{Int}$ and the total substitutional Te concentration. Such a strong correlation affirms that the electronic states introduced by substitutional Te are the origin of the sub-bandgap absorptance in hyperdoped Si. This is in great agreement with the previous work reported in Ref. 32 and 46, where the thermal anneals were also found to result in the deactivation of optical-active centers responsible for the sub-bandgap absorptance [29,47]. However, the Te-hyperdoped Si system exhibits a thermal stability and tolerates annealing up to 400 °C, which is promising for the integration of CMOS-compatible optoelectronics.

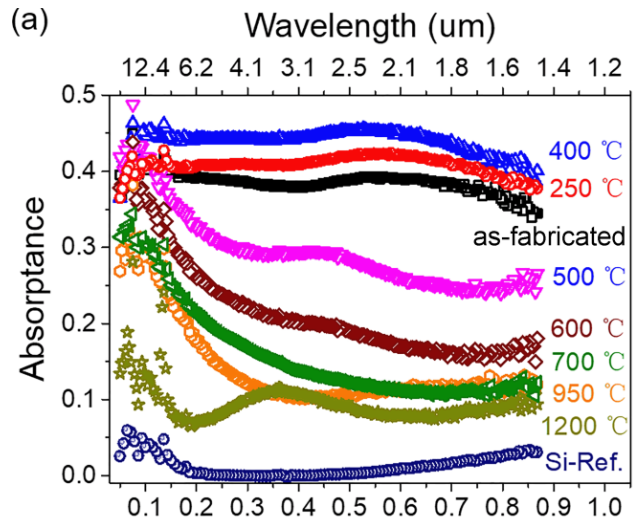
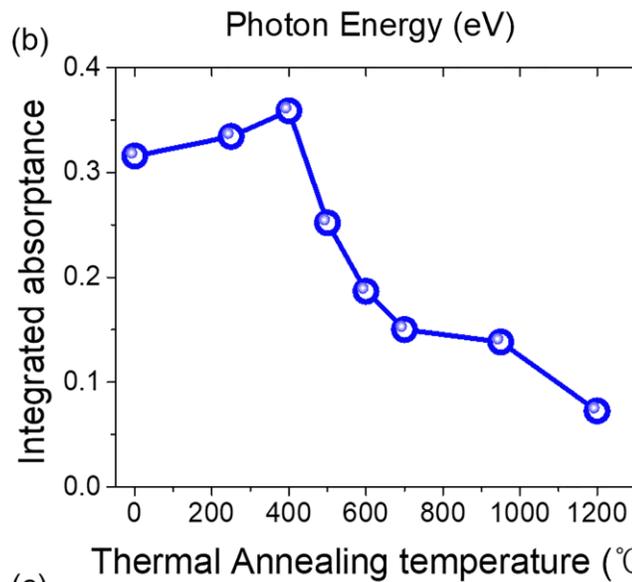
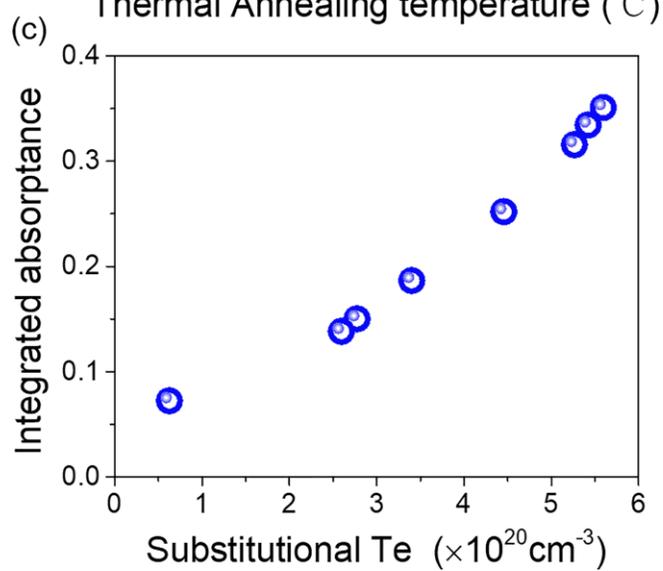

FIG. 6. (a) Sub-bandgap absorptance of as-fabricated Te-hyperdoped Si samples with an annealing duration of 10 min at different temperatures. The annealed temperatures are indicated as labels. The reference absorptance of an as-fabricated sample and the pure Si substrate is also included. (b) Integrated absorptance ($A_{Int}$) from 0.20 to 0.87 eV of the furnace annealed samples as a function of annealing temperature. (c) $A_{Int}$ of the as-fabricated sample and furnace annealed samples as a function of substitutional Te concentration (determined by RBS/C).

## IV. CONCLUSION

We have examined the thermal stability of the Te-hyperdoped Si subjected to isochronal anneals from 250 °C to 1200 °C. We have demonstrated that the recrystallization quality, the Te substitutionality, the electrical activation and the room-temperature sub-bandgap absorption exhibit further improvements upon furnace annealing at temperatures of 400 °C with a duration of 10 minutes. However, an impairment of the structural, electrical and optical properties is found as the annealing temperature increases above 400 °C. We have proven that the Te-hyperdoped Si layer possesses a thermal stability which is crucial for the realization of CMOS-compatible devices. In particular, we have found that the electrical and optical deactivation caused by the thermal treatment is attributed to the thermally-activated atomic-scale migration of Te dopants with migration energy of 2.1 ± 0.1 eV. At the atomic scale we have established a one-to-one correlation between the substitutional Te dopants and the unique properties. The knowledge of Te dopants transformation caused by thermal treatment contributes to achieve optimal properties such as an extremely high carrier concentration and strong sub-bandgap absorption for Si-based micro-/opto-electronics.

## ACKNOWLEDGMENTS

Authors acknowledge the Ion Beam Center (IBC) at Helmholtz-Zentrum Dresden-Rossendorf (HZDR) for performing the Te implantations. M. W. thanks Ilona Skorupa for her assistance in furnace annealing. Additionally, support by the Structural Characterization Facilities at Ion Beam Center (IBC) and funding of TEM Talos by the German Federal Ministry of Education of Research (BMBF), Grant No. 03SF0451 in the framework of HEMCP are gratefully acknowledged. This work is funded by the Helmholtz-Gemeinschaft Deutscher Forschungszentren (HGF-VH-NG-713). M.W. thanks financial support by Chinese Scholarship Council (File No. 201506240060). Y.B. would like to thank the Alexander-von-Humboldt foundation for providing a postdoctoral